\pdfoutput=1
\documentclass[aps,prl,twocolumn,groupedaddress]{revtex4-1}
\usepackage{graphicx}
\begin{document}   

%\preprint{}

\title{A Search for Dark Matter Axions with the Orpheus Experiment}


\author{Gray Rybka}
\email[]{grybka@uw.edu}
\affiliation{University of Washington}
\author{Andrew Wagner}
\email[]{apwagner@uw.edu}
\affiliation{University of Washington}
\author{Aryeh Brill}
\affiliation{Yale University}
\author{Kunal Patel}
\affiliation{University of Washington}
\author{Robert Percival}
\affiliation{University of Washington}
\author{Katleiah Ramos}
\affiliation{University of Washington}
%\thanks{}


\date{\today}

\begin{abstract}
Axions are well motivated particles that could make up most or all of the dark matter if they have masses below 100 $\mu$eV. Microwave cavity techniques comprised of closed resonant structures immersed in solenoid magnets are sensitive to dark matter axions with masses of a few~$\mu$eV, but face difficulties scaling to higher masses. We present the a novel detector architecture consisting of an open, Fabry-P\'{e}rot resonator and a series of current-carrying wire planes, and demonstrate this technique with a search for dark matter axion-like particles called Orpheus. This search excludes dark matter axion-like particles with masses between 68.2 and 76.5 $\mu$eV and axion-photon couplings greater than $4\times10^{-7}~\mathrm{GeV}^{-1}$. We project that the fundamental sensitivity of this technique could be extended to be sensitive to couplings below $1\times10^{-15}~\mathrm{GeV}^{-1}$, consistent with the DFSZ model of QCD axions. 
\end{abstract}

\pacs{}
%\keywords{}

\maketitle

\section{Introduction}

The axion is a pseudo-scalar particle predicted as a consequence to the Peccei-Quinn solution to the Strong CP problem~\cite{Peccei,Peccei_2,PhysRevLett.40.223,PhysRevLett.40.279}, and may comprise some or all of dark matter~\cite{Preskill1983127,Abbott1983,ipser-sikivie}. The axion has weak coupling to the electromagnetic interaction arising at loop order, whose Lagrange density may be written compactly as 

\begin{equation}
{\cal L}_{a\gamma\gamma}=-g_{a\gamma\gamma}a\vec{E}\cdot\vec{B}~,
\label{eqn:lagrangian}
\end{equation}

\noindent where $g_{a\gamma\gamma}$ is the axion-photon coupling strength, $a$ is the axion field, and $\vec{E}$,~$\vec{B}$ are the usual electric and magnetic fields. The expression in Eqn.~\ref{eqn:lagrangian} motivates the Axion Haloscope technique~\cite{PhysRevLett.51.1415} to detect dark matter axions. A typical Axion Haloscope consists of a closed microwave resonator immersed in a high static magnetic field, coupled to a low noise microwave receiver via the lowest frequency TM mode of the resonator.  Dark matter axions passing through the magnetic field can convert into photons inside the cavity with enhanced probability when an electromagnetic resonance in the cavity is tuned to correspond to the frequency of the photons produced. Dark matter axions would be detected as excess power at this frequency, the expression for which can be derived from Eqn.~\ref{eqn:lagrangian} as \cite{PhysRevLett.80.2043}

\begin{equation}
P=\frac{2\pi\hbar^2 g^2_{a\gamma\gamma}\rho_{\rm DM}}{m^2_ac}\cdot  f_{\gamma}\cdot \frac{1}{\mu_0}B^2V_{nlm}\cdot Q~.
\label{eqn:axpow}
\end{equation}

\noindent Here the $m_a$, $f_{\gamma}$ denote the axion mass and frequency of the converted photon respectively and $\rho_{\rm DM}\approx 0.4~{\rm GeV/cc}$ is the local halo density of dark matter. The enhancement in the expected axion power due to its conversion in a resonant cavity is expressed in terms of the cavity quality factor $Q$. The effective volume of the cavity for coupling to a given resonant mode is \cite{Peng2000569}

\begin{equation}
V_{nlm}=\frac{\left(\int d^3\vec{x} \vec{E}(\vec{x})\cdot\vec{B}(\vec{x})\right)^2}{B^2\int d^3\vec{x} |\vec{E}|^2(\vec{x})}~,
\label{eqn:Vnlm}
\end{equation}

\noindent where $\vec{B}(\vec{x})$ is the static magnetic field and $\vec{E}$ is the electric field of a normal resonant mode denoted by integers $n,~l,~m$. 

Numerous experiments based on this architecture have been constructed. Recently, the ADMX collaboration has demonstrated that microwave cavity experiments can be built with the sensitivity necessary to detect dark matter axions with masses in the range, 1.90--3.54 $\mu$eV~\cite{PhysRevLett.80.2043,PhysRevLett.104.041301} and coupling strength consistent with QCD predictions.  Some models, however, predict the axion mass scale to be somewhat larger~\cite{PhysRevD.80.035024, Khlopov1999105,PhysRevD.85.105020}.  Work is underway to extend experimental reach to larger axion masses, but the closed resonator detector design is difficult to extend to masses as large as 100~$\mu$eV~\cite{PhysRevD.64.092003}. Physically the size of a closed resonator must decrease in order to  achieve higher resonant frequencies. This in turn decreases both the volume and $Q$ of the resonator, which both limits the sensitivity of experiments based on this architecture and presents a serious challenge to their scalability. We present a dark matter axion search technique which overcomes the fundamental limitations of closed resonator architectures at large axion masses by employing an open, Fabry-P\'{e}rot resonator as the detector volume.  This technique is demonstrated by a prototype experiment named Orpheus.

\section{Detector}

\begin{figure}
\begin{center}
\includegraphics[width=9cm]{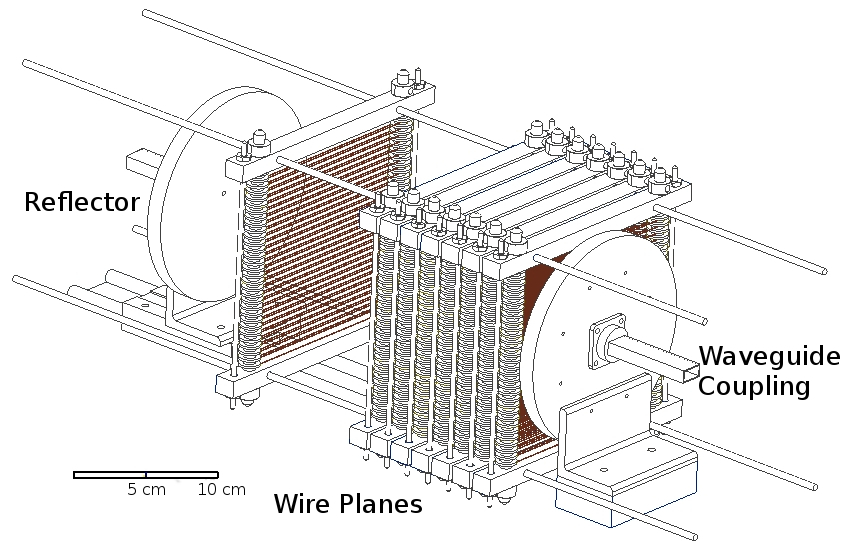}
\caption{\label{fig:cutaway} Model of the Orpheus experiment. The experiment consists of a regular grid of wires braced by a pair of reflectors forming a Fabry-P\'{e}rot resonator. The wire planes and reflectors are supported on rails permitting the frequency of the resonator to be adjusted with optimal alignment of the magnetic field supplied by the wires.}
\end{center}
\end{figure}

The Orpheus experiment consists of a half Fabry-P\'{e}rot resonator containing a series of wire planes (Fig.~\ref{fig:cutaway}) used to generate a static, spatially varying magnetic field, as described in~\cite{PhysRevD.50.4744}, within the volume of the detector.
Resonant modes with an electric field parallel to the magnetic field generated by the wire planes can couple to dark matter axions.  Comparing Eqn.~\ref{eqn:Vnlm}, to the transverse electromagnetic (TEM) modes of a Fabry-P\'{e}rot resonator~\cite{0022-3735-15-1-002}, it can be seen that the best coupling is achieved with the mode that has no nodes perpendicular to the axis of the resonator, and a node at every wire plane along the axis of the resonator, so the half-wavelength is close to the spacing between wire planes.  
This way the $\vec{E}(\vec{x})\cdot\vec{B}(\vec{x})$ integral over the volume has the same sign between each wire plane, and adds coherently.
We will refer to these modes as the ${\rm TEM}_{00-N}$ modes, where N is the number of nodes along the axis of the resonator.
The reflectors and wire planes are supported on independent rails, permitting the cavity and wire spacing to be adjusted and a search to be conducted over a range of axion masses. Power was coupled out of the detector volume via WR62 waveguide at frequencies around 17 GHz.
The mode with a half wavelength closest to the wire plane separation, given our reflector geometry and separation, corresponded to the ${\rm TEM}_{00-19}$ mode of the resonator. 
 The experiment was operated at room temperature, and the axion search bandwidth was approximately 2 GHz.

\begin{figure}
\begin{center}
\includegraphics[width=9cm]{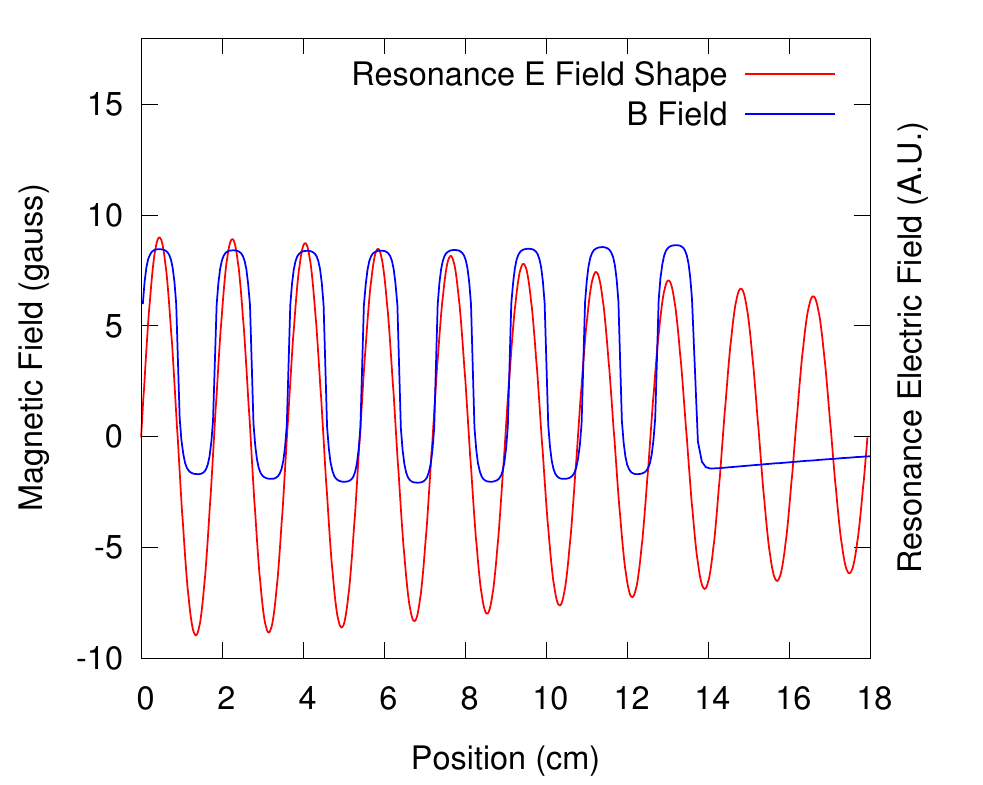}
\caption{\label{fig:EMfields} Simulated magnetic and electric fields on-axis within the Fabry-P\'{e}rot resonator.  Both fields are primarily in the vertical direction relative to Fig.~\ref{fig:cutaway}, perpendicular to the wires in the wire planes.   The maximum magnetic field for this experiment was 8.5 gauss, the electric field of the ${\rm TEM}_{00-19}$ mode is shown on same scale in arbitrary units for clarity.}
\end{center}
\end{figure}

The half Fabry-P\'{e}rot resonator consisted of two aluminum reflectors, one flat and a second with a radius of curvature of 33 cm, each 15 cm in diameter. Critical coupling was achieved via a waveguide stub tuner immediately following an aperture to waveguide adapter on the flat reflector. A similar aperture existed on the curved reflector permitting weak coupling of power into the resonator. The unloaded $Q$ of the ${\rm TEM}_{00-19}$ mode was 4,500 when tuned to 17 GHz, and 11,600 at 18 GHz. The magnetic field was generated by 8 wire frames, constructed from 0.32 mm diameter copper wire wound in a plane on frames 16 cm by 15 cm. Each frame contained two wire planes. The spacing between adjacent wires on a single plane was 4.9 mm, and the spacing between planes on a single frame was 9.1 mm. The interframe spacing was adjusted by sliding the wire frames along their support rails. When energized with 3.4 A, the field within a frame was 8.5 G and the field between frames was 3.5 G, in close agreement with simulation (Fig.~\ref{fig:EMfields}). Optimal placement of the wire frames inside the resonator was achieved by adjusting their position so as to maximize the $Q$ of the ${\rm TEM}_{00-19}$ mode. This occurs when the planes are placed at the nodes of the EM mode within the resonator. The optimized, loaded $Q$ of the resonator containing the wire planes was 4,300 at 17 GHz and 7,900 at 18 GHz.

\begin{figure}[h]
\begin{center}
\includegraphics[width=9cm]{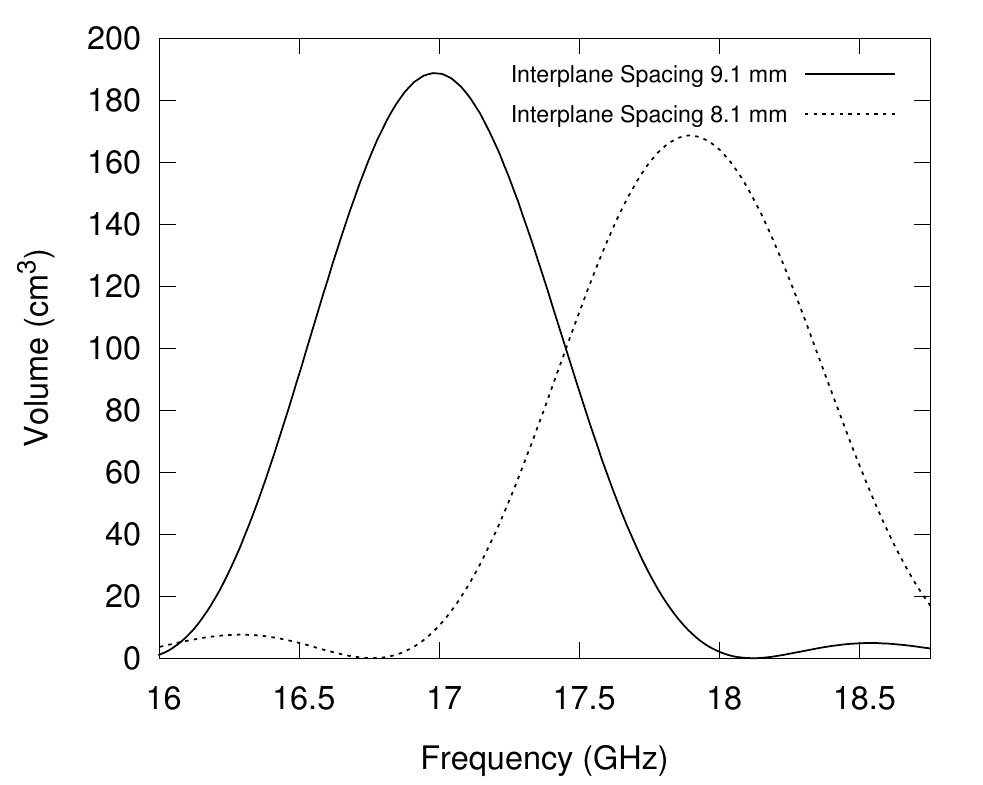}
\caption{\label{fig:Vnlm} $V_{nlm}$ as defined in Eqn.~\ref{eqn:Vnlm} for two different wire frame spacings. 9.1 mm spacing (solid) 8.1 mm spacing (dashed)}
\end{center}
\end{figure}

Fig.~\ref{fig:EMfields} illustrates the electric field of the ${\rm TEM}_{00-19}$ mode in the Fabry-P\'{e}rot resonator and the static magnetic field provided by the wire planes. The resonance field profile is Gaussian in the radial direction~\cite{0022-3735-15-1-002}. We numerically evaluate Eqn.~\ref{eqn:Vnlm} using the magnetic and electric field profiles in Fig.~\ref{fig:EMfields} and plot $V_{nlm}$ as a function of frequency for the two reflector separations in Fig.~\ref{fig:Vnlm}. When the magnetic field is aligned with the resonance, $V_{nlm}$ is roughly the beam cross section at the center of the resonator multiplied by the distance between the reflectors, independent of the resonant frequency. Unlike closed cavity experiments, the volume of this setup can be easily increased by moving the reflectors further apart and adding more wire planes.

%The power $P$ in watts of axion-to-photon conversion in the resonator (assuming the KSVZ model of axion coupling\cite{Kim19871}) can be expressed in a manner similar to that for resonant cavity detectors given in Ref.~\cite{Cavity_idea_2}.

%\begin{widetext}
%\begin{equation}
%P=2.2\times10^{-23} W \left(\frac{V_{\mathrm{eff}}}{1000\ \mathrm{cm^{3}}}\right)\left(\frac{B}{1\ \mathrm{T}}\right)^2\left(\frac{\rho}{0.45\ \mathrm{GeV cm^{-3}}}\right)\left(\frac{f}{100\ \mathrm{GHz}}\right)\left(\frac{\mathrm{min}(Q,10^6)}{100,000}\right)
%\label{eqn:powereqn}
%\end{equation}
%\end{widetext}

\section{Results and Analysis}

The procedure for conducting a search for dark matter axions with Orpheus proceeds in a manner analogous to that used in closed resonator architectures such as ADMX~\cite{PhysRevLett.80.2043,PhysRevLett.104.041301}.  First, a network analyzer (HP Model-8753C) was used to locate the frequency of the desired resonance and determine the Q.  The amplified power spectral density across the resonant bandwidth was then recorded with a spectrum analyzer (Agilent Model-N9000A). Following the power measurement, a stepper motor was used to adjust the distance between the reflectors to change the resonant frequency by half the resonance bandwidth. This process was repeated and the effective volume given in Eq.~\ref{eqn:Vnlm} is shown in Fig.~\ref{fig:Vnlm} over the range of frequencies scanned. 

\begin{figure}
\begin{center}
\includegraphics[width=9cm]{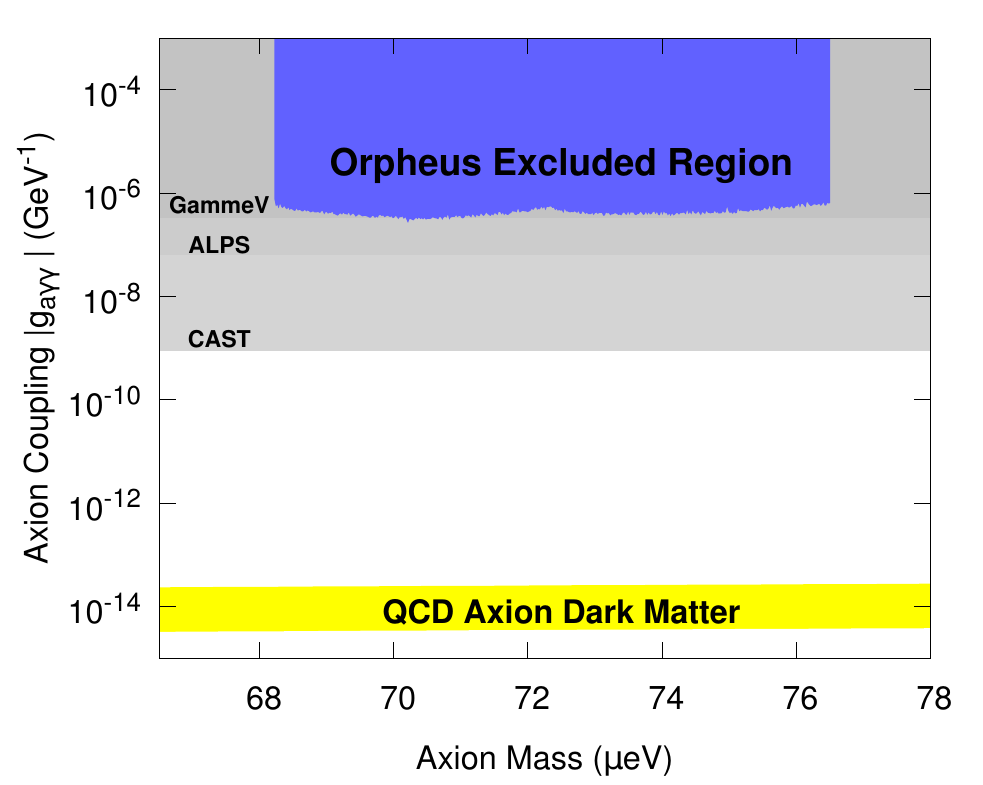}
\caption{\label{fig:thelimit} Axion-photon coupling $g_{a\gamma\gamma}$, and masses excluded from this experiment, assuming axion-like particles make up all of the local dark matter density. Limits from the laser-based experiments GammeV\cite{PhysRevLett.100.080402} and ALPS\cite{Ehret2010149}, and the solar experiment CAST\cite{PhysRevLett.107.261302} are shown for comparison.}
\end{center}
\end{figure}

An axion signal would appear as a peak in the recorded power spectra within a bandwidth of 16.5 kHz, (the thermal broadening assuming a virialized local dark matter distribution \cite{PhysRevD.42.3572}) and significance $S$ given by

\begin{equation}
S=\frac{P}{k_BT}\sqrt{\frac{t}{b}}~,
\label{eqn:nyquist}
\end{equation}

\noindent where $P$ is the expected signal power given by Eqn.~\ref{eqn:axpow}, $t$ is the signal integration time, $b$ the signal bandwidth and $T$ the system noise temperature. During this experiment, the spectral density was recorded over a bandwidth of roughly 6.7 MHz, integrating for 4.5 seconds and binned into 16.5 kHz bins, and the typical tuning resonant frequency step size was 3 MHz.  The frequency range 16.5 GHz to 18.5 GHz was covered in roughly a week of running.  The sensitivity is limited by the noise temperature of the system, which comes from the combination of the noise temperature of the room temperature commercial amplifiers and the physical temperature of the resonator.  From these, allowing for some cable loss, we use the conservative estimate of a system noise of 1420 K.  In order to combine individual spectra to produce a limit on the axion-photon coupling, $g_{a\gamma\gamma}$,  power spectra were fit with a 5th order polynomial to remove the broadband receiver transfer function and normalized to the measured system noise temperature. The calibrated spectra were then added and used to calculate the expected value of the axion-photon coupling required to produce any excess power in a given bin, assuming an axion conversion signal was present.  An axion signal would appear as a narrow excess of power in the otherwise flat background-subtracted power spectra.  The spectra were statistically consistent with no axion signal, and subsequently a 95\% confidence limit is set on the axion-photon coupling over the frequency range covered during the experiments operation. The result of this search excludes dark matter axion-like particle masses between 68.2 and 76.5 $\mu$eV with axion-photon couplings greater than $4\times10^{-7}~\mathrm{GeV}^{-1}$ as shown in Fig.~\ref{fig:thelimit}.

\section{Achievable Sensitivity of Technique}

\begin{table*}
\begin{tabular}{|c|c|c|c|c|c|c|c|}
\hline
Experiment & Mass Target    & Frequency & B Field & Q        & Volume  &  Noise Temperature & Run Time\\
\hline
A          &     52 $\mu$eV & 15 GHz     & 3 T & $10^{6}$ &   $1\times10^{6}\ \mathrm{cm}^3$  &    750 mK & 1 Year\\
B          &     103 $\mu$eV & 30 GHz      & 3 T & $10^{6}$ &   $8\times10^{5}\ \mathrm{cm}^3$  &    1.5 K & 1 Year\\
C          &     207 $\mu$eV & 60 GHz      & 6 T & $10^{6}$ &   $4\times10^{5}\ \mathrm{cm}^3$  &    3 K & 1 Year\\
D          &     414 $\mu$eV & 120 GHz     & 6 T & $10^{6}$ &   $2\times10^{5}\ \mathrm{cm}^3$  &    6 K & 1 Year\\
\hline
\end{tabular}
\caption{\label{tab:expparams}Estimated parameters for axion experiments.}
\end{table*}

\begin{figure}[h]
\includegraphics[width=8cm]{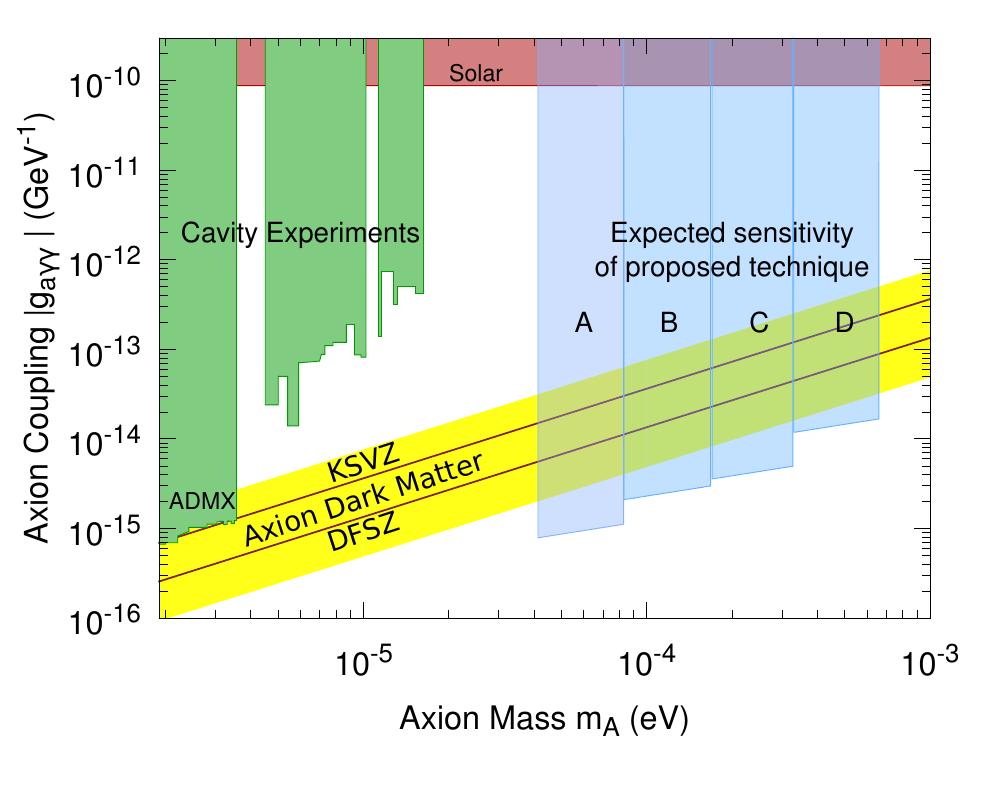}
\caption{\label{fig:projected_sensitivity} Expected sensitivity to axion-photon coupling of proposed experiment technique for the experimental parameters listed in table~\ref{tab:expparams} assuming axions make up all of the local dark matter density.
Also shown for comparison are limits set by existing microwave cavity experiments \cite{PhysRevLett.80.2043,PhysRevLett.104.041301,PhysRevLett.59.839,PhysRevD.40.3153,PhysRevD.42.1297}, and solar axion experiments\cite{PhysRevLett.107.261302}.  The ADMX experiment is currently being upgraded to expand its sensitivity to 40~$\mu$eV at DFSZ sensitivity, making the proposed technique a natural extension of the search range. \cite{admxpatras2013}}
\end{figure}

It is important to note that the sensitivity demonstrated in Fig.~\ref{fig:thelimit} reflects the small scale implementation of the current incarnation of the Orpheus experiment and does not indicate a fundamental limitation of the technique. The ultimate sensitivity is determined by the technology used to produce the magnetic field, the quality factor and volume of the resonator, and the noise temperature of the receiving electronics as indicated by Eqns.~\ref{eqn:axpow}~and~\ref{eqn:nyquist}. We now discuss the achievable values of these parameters, and project the reach of experiments using this technique and show that it can be sensitive to even a pessimistically coupled axion dark matter, or dark matter scenarios where axions do not make up all of the dark matter. Our estimates based on relevant experimental parameters are summarized for a number of hypothetical experiments in several frequency ranges in Tab.~\ref{tab:expparams}. The potential reach of such experiments is shown in Fig.~\ref{fig:projected_sensitivity}. 

The magnetic field was limited in this experiment by the finite conductivity of the wire used to construct the current carrying planes.  For an optimal experiment, the magnetic fields used should be as large as possible.  Greatly increased magnetic fields can be achieved by constructing the planes from superconducting wires. We estimate that planes of wires with 0.4 mm spacing carrying 470 A could support a 3 T field. Power supplies and superconducting NbTi wire capable of supporting these currents are commercially available. The generation of higher magnetic fields will require finer wire spacing for the same supplied current. Such small spacing is impractical to achieve with wound superconducting wire but might be possible by employing thick film micro-fabrication techniques. The critical current density of $\rm{Nb}_3\rm{Sn}$ films have been shown to be as high $4\times10^4 \rm{A}/\rm{mm}^2$ at 6 T~\cite{Nb3Sn}. This field could be supported by photolithographically patterned wires with 60 $\mu$m spacing carrying 144 A. 

With appropriate wire spacing and the addition of field cancellation planes it may be possible to eliminate the residual magnetic field within a few wavelengths of the resonant mode. This allows the possibility of using superconducting reflectors in the construction of the Fabry-P\'{e}rot resonator~\cite{:/content/aip/journal/apl/90/16/10.1063/1.2724816}, making an all superconducting implementation of the architecture and attractive possibility. Such superconducting resonators have demonstrated $Q$ values of $5\times10^{9}$. The intervening wire planes will decrease this somewhat by diffraction, so careful design of the wire chamber and reflector will need to maximize the product $B^2V_{nlm}\cdot Q$. 

The resonator length is limited fundamentally by the axion wavelength, and practically by beam alignment, reflector fabrication and the feasibility of maintaining a large volume at cryogenic temperatures.
%The axion wavelength is given by 
%\begin{equation}
%\lambda=\frac{h}{m_a\beta c}\sqrt{1-\beta^2}~,
%\label{eqn:deBrolgie}
%\end{equation}
%\noindent where $\beta\approx 0.73\times 10^{-3}$ for virialized cold dark matter. We estimate the longest convenient detector is thus $\approx 10$ m for axion masses up to 170~$\mu$eV. 
As dark matter axions are expected to be nonrelativistic, they have long wavelengths that allow coherent interaction with detectors as long as 10 m for axion masses up to 170~$\mu$eV.  We therefore assume detectors with a length scale of 10 m.
The maximum beam width will depend on the fabrication of the reflectors, and given current fabrication technologies we estimate the beam width will scale roughly as $(50\ \mathrm{cm}) \sqrt{\frac{10 \mathrm{GHz}}{f}}$.

Finally we consider the improvement that implementing quantum noise limited superconducting amplifiers would bring to a cryogenic experiment. Josephson Parametric Amplifiers (JPAs) have been operated as phase preserving amplifiers with noise temperatures near the quantum limit, roughly 50 $\frac{\mathrm{mK}}{\mathrm{GHz}}$, at frequencies near 19 GHz~\cite{Yurke:1996dt}. While the uncertainty principle places a limit on sensitivity with which a power spectrum can be measured, this restriction can be evaded provided one only cares to measure the occurrence of an axion-photon conversion and not its spectral density. Single photon counting techniques have been demonstrated based on superconducting Josephson qubits at 4 GHz~\cite{Chen:2011fj} and might be engineered to operate at higher frequencies to the benefit of axion experiments \cite{PhysRevD.88.035020}.

\section{Conclusions}

We have presented a technique applicable to dark matter axion searches in the mass range 40--400 $\mu$eV, and demonstrated the technique with an experiment that searches for axion-like particles in the 68.2--76.5 $\mu$eV mass range, which is favored by some models of axion dark matter.\cite{PhysRevD.80.035024}  Reasonable estimates based on the technology available, combined with the performance of a small scale implementation, suggest that experiments using this technique could be constructed to explore the majority of theoretically allowed axion-photon couplings.

This work was supported in part by the U.S. Department of Energy under contract DE-SC0009800.

\bibliography{2014_orpheus}

\begin{thebibliography}{30}%
\makeatletter
\providecommand \@ifxundefined [1]{%
 \@ifx{#1\undefined}
}%
\providecommand \@ifnum [1]{%
 \ifnum #1\expandafter \@firstoftwo
 \else \expandafter \@secondoftwo
 \fi
}%
\providecommand \@ifx [1]{%
 \ifx #1\expandafter \@firstoftwo
 \else \expandafter \@secondoftwo
 \fi
}%
\providecommand \natexlab [1]{#1}%
\providecommand \enquote  [1]{``#1''}%
\providecommand \bibnamefont  [1]{#1}%
\providecommand \bibfnamefont [1]{#1}%
\providecommand \citenamefont [1]{#1}%
\providecommand \href@noop [0]{\@secondoftwo}%
\providecommand \href [0]{\begingroup \@sanitize@url \@href}%
\providecommand \@href[1]{\@@startlink{#1}\@@href}%
\providecommand \@@href[1]{\endgroup#1\@@endlink}%
\providecommand \@sanitize@url [0]{\catcode `\\12\catcode `\$12\catcode
  `\&12\catcode `\#12\catcode `\^12\catcode `\_12\catcode `\%12\relax}%
\providecommand \@@startlink[1]{}%
\providecommand \@@endlink[0]{}%
\providecommand \url  [0]{\begingroup\@sanitize@url \@url }%
\providecommand \@url [1]{\endgroup\@href {#1}{\urlprefix }}%
\providecommand \urlprefix  [0]{URL }%
\providecommand \Eprint [0]{\href }%
\providecommand \doibase [0]{http://dx.doi.org/}%
\providecommand \selectlanguage [0]{\@gobble}%
\providecommand \bibinfo  [0]{\@secondoftwo}%
\providecommand \bibfield  [0]{\@secondoftwo}%
\providecommand \translation [1]{[#1]}%
\providecommand \BibitemOpen [0]{}%
\providecommand \bibitemStop [0]{}%
\providecommand \bibitemNoStop [0]{.\EOS\space}%
\providecommand \EOS [0]{\spacefactor3000\relax}%
\providecommand \BibitemShut  [1]{\csname bibitem#1\endcsname}%
\let\auto@bib@innerbib\@empty
%</preamble>
\bibitem [{\citenamefont {Peccei}\ and\ \citenamefont
  {Quinn}(1977{\natexlab{a}})}]{Peccei}%
  \BibitemOpen
  \bibfield  {author} {\bibinfo {author} {\bibfnamefont {R.}~\bibnamefont
  {Peccei}}\ and\ \bibinfo {author} {\bibfnamefont {H.}~\bibnamefont {Quinn}},\
  }\href@noop {} {\bibfield  {journal} {\bibinfo  {journal} {Phys. Rev. Lett.}\
  }\textbf {\bibinfo {volume} {38}},\ \bibinfo {pages} {1440} (\bibinfo {year}
  {1977}{\natexlab{a}})}\BibitemShut {NoStop}%
\bibitem [{\citenamefont {Peccei}\ and\ \citenamefont
  {Quinn}(1977{\natexlab{b}})}]{Peccei_2}%
  \BibitemOpen
  \bibfield  {author} {\bibinfo {author} {\bibfnamefont {R.}~\bibnamefont
  {Peccei}}\ and\ \bibinfo {author} {\bibfnamefont {H.}~\bibnamefont {Quinn}},\
  }\href@noop {} {\bibfield  {journal} {\bibinfo  {journal} {Phys. Rev. D}\
  }\textbf {\bibinfo {volume} {16}},\ \bibinfo {pages} {1791} (\bibinfo {year}
  {1977}{\natexlab{b}})}\BibitemShut {NoStop}%
\bibitem [{\citenamefont {Weinberg}(1978)}]{PhysRevLett.40.223}%
  \BibitemOpen
  \bibfield  {author} {\bibinfo {author} {\bibfnamefont {S.}~\bibnamefont
  {Weinberg}},\ }\href {\doibase 10.1103/PhysRevLett.40.223} {\bibfield
  {journal} {\bibinfo  {journal} {Phys. Rev. Lett.}\ }\textbf {\bibinfo
  {volume} {40}},\ \bibinfo {pages} {223} (\bibinfo {year} {1978})}\BibitemShut
  {NoStop}%
\bibitem [{\citenamefont {Wilczek}(1978)}]{PhysRevLett.40.279}%
  \BibitemOpen
  \bibfield  {author} {\bibinfo {author} {\bibfnamefont {F.}~\bibnamefont
  {Wilczek}},\ }\href {\doibase 10.1103/PhysRevLett.40.279} {\bibfield
  {journal} {\bibinfo  {journal} {Phys. Rev. Lett.}\ }\textbf {\bibinfo
  {volume} {40}},\ \bibinfo {pages} {279} (\bibinfo {year} {1978})}\BibitemShut
  {NoStop}%
\bibitem [{\citenamefont {Preskill}\ \emph {et~al.}(1983)\citenamefont
  {Preskill}, \citenamefont {Wise},\ and\ \citenamefont
  {Wilczek}}]{Preskill1983127}%
  \BibitemOpen
  \bibfield  {author} {\bibinfo {author} {\bibfnamefont {J.}~\bibnamefont
  {Preskill}}, \bibinfo {author} {\bibfnamefont {M.}~\bibnamefont {Wise}}, \
  and\ \bibinfo {author} {\bibfnamefont {F.}~\bibnamefont {Wilczek}},\ }\href
  {\doibase 10.1016/0370-2693(83)90637-8} {\bibfield  {journal} {\bibinfo
  {journal} {Phys. Lett. B}\ }\textbf {\bibinfo {volume} {120}},\ \bibinfo
  {pages} {127} (\bibinfo {year} {1983})}\BibitemShut {NoStop}%
\bibitem [{\citenamefont {Abbott}\ and\ \citenamefont
  {Sikivie}(1983)}]{Abbott1983}%
  \BibitemOpen
  \bibfield  {author} {\bibinfo {author} {\bibfnamefont {L.}~\bibnamefont
  {Abbott}}\ and\ \bibinfo {author} {\bibfnamefont {P.}~\bibnamefont
  {Sikivie}},\ }\href@noop {} {\bibfield  {journal} {\bibinfo  {journal} {Phys.
  Lett. B}\ }\textbf {\bibinfo {volume} {120}},\ \bibinfo {pages} {133}
  (\bibinfo {year} {1983})}\BibitemShut {NoStop}%
\bibitem [{\citenamefont {Ipser}\ and\ \citenamefont
  {Sikivie}(1983)}]{ipser-sikivie}%
  \BibitemOpen
  \bibfield  {author} {\bibinfo {author} {\bibfnamefont {J.}~\bibnamefont
  {Ipser}}\ and\ \bibinfo {author} {\bibfnamefont {P.}~\bibnamefont
  {Sikivie}},\ }\href {\doibase 10.1103/PhysRevLett.50.925} {\bibfield
  {journal} {\bibinfo  {journal} {Phys. Rev. Lett.}\ }\textbf {\bibinfo
  {volume} {50}},\ \bibinfo {pages} {925} (\bibinfo {year} {1983})}\BibitemShut
  {NoStop}%
\bibitem [{\citenamefont {Sikivie}(1983)}]{PhysRevLett.51.1415}%
  \BibitemOpen
  \bibfield  {author} {\bibinfo {author} {\bibfnamefont {P.}~\bibnamefont
  {Sikivie}},\ }\href {\doibase 10.1103/PhysRevLett.51.1415} {\bibfield
  {journal} {\bibinfo  {journal} {Phys. Rev. Lett.}\ }\textbf {\bibinfo
  {volume} {51}},\ \bibinfo {pages} {1415} (\bibinfo {year}
  {1983})}\BibitemShut {NoStop}%
\bibitem [{\citenamefont {Hagmann}\ \emph {et~al.}(1998)\citenamefont
  {Hagmann}, \citenamefont {Kinion}, \citenamefont {Stoeffl}, \citenamefont
  {van Bibber}, \citenamefont {Daw}, \citenamefont {Peng}, \citenamefont
  {Rosenberg}, \citenamefont {LaVeigne}, \citenamefont {Sikivie}, \citenamefont
  {Sullivan} \emph {et~al.}}]{PhysRevLett.80.2043}%
  \BibitemOpen
  \bibfield  {author} {\bibinfo {author} {\bibfnamefont {C.}~\bibnamefont
  {Hagmann}}, \bibinfo {author} {\bibfnamefont {D.}~\bibnamefont {Kinion}},
  \bibinfo {author} {\bibfnamefont {W.}~\bibnamefont {Stoeffl}}, \bibinfo
  {author} {\bibfnamefont {K.}~\bibnamefont {van Bibber}}, \bibinfo {author}
  {\bibfnamefont {E.}~\bibnamefont {Daw}}, \bibinfo {author} {\bibfnamefont
  {H.}~\bibnamefont {Peng}}, \bibinfo {author} {\bibfnamefont {L.~J.}\
  \bibnamefont {Rosenberg}}, \bibinfo {author} {\bibfnamefont {J.}~\bibnamefont
  {LaVeigne}}, \bibinfo {author} {\bibfnamefont {P.}~\bibnamefont {Sikivie}},
  \bibinfo {author} {\bibfnamefont {N.~S.}\ \bibnamefont {Sullivan}},  \emph
  {et~al.},\ }\href {\doibase 10.1103/PhysRevLett.80.2043} {\bibfield
  {journal} {\bibinfo  {journal} {Phys. Rev. Lett.}\ }\textbf {\bibinfo
  {volume} {80}},\ \bibinfo {pages} {2043} (\bibinfo {year}
  {1998})}\BibitemShut {NoStop}%
\bibitem [{\citenamefont {Peng}\ \emph {et~al.}(2000)\citenamefont {Peng},
  \citenamefont {Asztalos}, \citenamefont {Daw}, \citenamefont {Golubev},
  \citenamefont {Hagmann}, \citenamefont {Kinion}, \citenamefont {LaVeigne},
  \citenamefont {Moltz}, \citenamefont {Nezrick}, \citenamefont {Powell} \emph
  {et~al.}}]{Peng2000569}%
  \BibitemOpen
  \bibfield  {author} {\bibinfo {author} {\bibfnamefont {H.}~\bibnamefont
  {Peng}}, \bibinfo {author} {\bibfnamefont {S.}~\bibnamefont {Asztalos}},
  \bibinfo {author} {\bibfnamefont {E.}~\bibnamefont {Daw}}, \bibinfo {author}
  {\bibfnamefont {N.}~\bibnamefont {Golubev}}, \bibinfo {author} {\bibfnamefont
  {C.}~\bibnamefont {Hagmann}}, \bibinfo {author} {\bibfnamefont
  {D.}~\bibnamefont {Kinion}}, \bibinfo {author} {\bibfnamefont
  {J.}~\bibnamefont {LaVeigne}}, \bibinfo {author} {\bibfnamefont
  {D.}~\bibnamefont {Moltz}}, \bibinfo {author} {\bibfnamefont
  {F.}~\bibnamefont {Nezrick}}, \bibinfo {author} {\bibfnamefont
  {J.}~\bibnamefont {Powell}},  \emph {et~al.},\ }\href {\doibase
  http://dx.doi.org/10.1016/S0168-9002(99)00971-7} {\bibfield  {journal}
  {\bibinfo  {journal} {Nuclear Instruments and Methods in Physics Research
  Section A: Accelerators, Spectrometers, Detectors and Associated Equipment}\
  }\textbf {\bibinfo {volume} {444}},\ \bibinfo {pages} {569 } (\bibinfo {year}
  {2000})}\BibitemShut {NoStop}%
\bibitem [{\citenamefont {Asztalos}\ \emph {et~al.}(2010)\citenamefont
  {Asztalos}, \citenamefont {Carosi}, \citenamefont {Hagmann}, \citenamefont
  {Kinion}, \citenamefont {van Bibber}, \citenamefont {Hotz}, \citenamefont
  {Rosenberg}, \citenamefont {Rybka}, \citenamefont {Hoskins}, \citenamefont
  {Hwang} \emph {et~al.}}]{PhysRevLett.104.041301}%
  \BibitemOpen
  \bibfield  {author} {\bibinfo {author} {\bibfnamefont {S.~J.}\ \bibnamefont
  {Asztalos}}, \bibinfo {author} {\bibfnamefont {G.}~\bibnamefont {Carosi}},
  \bibinfo {author} {\bibfnamefont {C.}~\bibnamefont {Hagmann}}, \bibinfo
  {author} {\bibfnamefont {D.}~\bibnamefont {Kinion}}, \bibinfo {author}
  {\bibfnamefont {K.}~\bibnamefont {van Bibber}}, \bibinfo {author}
  {\bibfnamefont {M.}~\bibnamefont {Hotz}}, \bibinfo {author} {\bibfnamefont
  {L.~J.}\ \bibnamefont {Rosenberg}}, \bibinfo {author} {\bibfnamefont
  {G.}~\bibnamefont {Rybka}}, \bibinfo {author} {\bibfnamefont
  {J.}~\bibnamefont {Hoskins}}, \bibinfo {author} {\bibfnamefont
  {J.}~\bibnamefont {Hwang}},  \emph {et~al.},\ }\href {\doibase
  10.1103/PhysRevLett.104.041301} {\bibfield  {journal} {\bibinfo  {journal}
  {Phys. Rev. Lett.}\ }\textbf {\bibinfo {volume} {104}},\ \bibinfo {pages}
  {041301} (\bibinfo {year} {2010})}\BibitemShut {NoStop}%
\bibitem [{\citenamefont {Visinelli}\ and\ \citenamefont
  {Gondolo}(2009)}]{PhysRevD.80.035024}%
  \BibitemOpen
  \bibfield  {author} {\bibinfo {author} {\bibfnamefont {L.}~\bibnamefont
  {Visinelli}}\ and\ \bibinfo {author} {\bibfnamefont {P.}~\bibnamefont
  {Gondolo}},\ }\href {\doibase 10.1103/PhysRevD.80.035024} {\bibfield
  {journal} {\bibinfo  {journal} {Phys. Rev. D}\ }\textbf {\bibinfo {volume}
  {80}},\ \bibinfo {pages} {035024} (\bibinfo {year} {2009})}\BibitemShut
  {NoStop}%
\bibitem [{\citenamefont {Khlopov}\ \emph {et~al.}(1999)\citenamefont
  {Khlopov}, \citenamefont {Sakharov},\ and\ \citenamefont
  {Sokoloff}}]{Khlopov1999105}%
  \BibitemOpen
  \bibfield  {author} {\bibinfo {author} {\bibfnamefont {M.}~\bibnamefont
  {Khlopov}}, \bibinfo {author} {\bibfnamefont {A.}~\bibnamefont {Sakharov}}, \
  and\ \bibinfo {author} {\bibfnamefont {D.}~\bibnamefont {Sokoloff}},\ }\href
  {\doibase http://dx.doi.org/10.1016/S0920-5632(98)00511-8} {\bibfield
  {journal} {\bibinfo  {journal} {Nuclear Physics B - Proceedings Supplements}\
  }\textbf {\bibinfo {volume} {72}},\ \bibinfo {pages} {105 } (\bibinfo {year}
  {1999})},\ \bibinfo {note} {proceedings of the 5th \{IFT\} Workshop on
  Axions}\BibitemShut {NoStop}%
\bibitem [{\citenamefont {Hiramatsu}\ \emph {et~al.}(2012)\citenamefont
  {Hiramatsu}, \citenamefont {Kawasaki}, \citenamefont {Saikawa},\ and\
  \citenamefont {Sekiguchi}}]{PhysRevD.85.105020}%
  \BibitemOpen
  \bibfield  {author} {\bibinfo {author} {\bibfnamefont {T.}~\bibnamefont
  {Hiramatsu}}, \bibinfo {author} {\bibfnamefont {M.}~\bibnamefont {Kawasaki}},
  \bibinfo {author} {\bibfnamefont {K.}~\bibnamefont {Saikawa}}, \ and\
  \bibinfo {author} {\bibfnamefont {T.}~\bibnamefont {Sekiguchi}},\ }\href
  {\doibase 10.1103/PhysRevD.85.105020} {\bibfield  {journal} {\bibinfo
  {journal} {Phys. Rev. D}\ }\textbf {\bibinfo {volume} {85}},\ \bibinfo
  {pages} {105020} (\bibinfo {year} {2012})}\BibitemShut {NoStop}%
\bibitem [{\citenamefont {Asztalos}\ \emph {et~al.}(2001)\citenamefont
  {Asztalos}, \citenamefont {Daw}, \citenamefont {Peng}, \citenamefont
  {Rosenberg}, \citenamefont {Hagmann}, \citenamefont {Kinion}, \citenamefont
  {Stoeffl}, \citenamefont {van Bibber}, \citenamefont {Sikivie}, \citenamefont
  {Sullivan} \emph {et~al.}}]{PhysRevD.64.092003}%
  \BibitemOpen
  \bibfield  {author} {\bibinfo {author} {\bibfnamefont {S.}~\bibnamefont
  {Asztalos}}, \bibinfo {author} {\bibfnamefont {E.}~\bibnamefont {Daw}},
  \bibinfo {author} {\bibfnamefont {H.}~\bibnamefont {Peng}}, \bibinfo {author}
  {\bibfnamefont {L.~J.}\ \bibnamefont {Rosenberg}}, \bibinfo {author}
  {\bibfnamefont {C.}~\bibnamefont {Hagmann}}, \bibinfo {author} {\bibfnamefont
  {D.}~\bibnamefont {Kinion}}, \bibinfo {author} {\bibfnamefont
  {W.}~\bibnamefont {Stoeffl}}, \bibinfo {author} {\bibfnamefont
  {K.}~\bibnamefont {van Bibber}}, \bibinfo {author} {\bibfnamefont
  {P.}~\bibnamefont {Sikivie}}, \bibinfo {author} {\bibfnamefont {N.~S.}\
  \bibnamefont {Sullivan}},  \emph {et~al.},\ }\href {\doibase
  10.1103/PhysRevD.64.092003} {\bibfield  {journal} {\bibinfo  {journal} {Phys.
  Rev. D}\ }\textbf {\bibinfo {volume} {64}},\ \bibinfo {pages} {092003}
  (\bibinfo {year} {2001})}\BibitemShut {NoStop}%
\bibitem [{\citenamefont {Sikivie}\ \emph {et~al.}(1994)\citenamefont
  {Sikivie}, \citenamefont {Tanner},\ and\ \citenamefont
  {Wang}}]{PhysRevD.50.4744}%
  \BibitemOpen
  \bibfield  {author} {\bibinfo {author} {\bibfnamefont {P.}~\bibnamefont
  {Sikivie}}, \bibinfo {author} {\bibfnamefont {D.~B.}\ \bibnamefont {Tanner}},
  \ and\ \bibinfo {author} {\bibfnamefont {Y.}~\bibnamefont {Wang}},\ }\href
  {\doibase 10.1103/PhysRevD.50.4744} {\bibfield  {journal} {\bibinfo
  {journal} {Phys. Rev. D}\ }\textbf {\bibinfo {volume} {50}},\ \bibinfo
  {pages} {4744} (\bibinfo {year} {1994})}\BibitemShut {NoStop}%
\bibitem [{\citenamefont {Clarke}\ and\ \citenamefont
  {Rosenberg}(1982)}]{0022-3735-15-1-002}%
  \BibitemOpen
  \bibfield  {author} {\bibinfo {author} {\bibfnamefont {R.~N.}\ \bibnamefont
  {Clarke}}\ and\ \bibinfo {author} {\bibfnamefont {C.~B.}\ \bibnamefont
  {Rosenberg}},\ }\href {http://stacks.iop.org/0022-3735/15/i=1/a=002}
  {\bibfield  {journal} {\bibinfo  {journal} {J. Phys. E}\ }\textbf {\bibinfo
  {volume} {15}},\ \bibinfo {pages} {9} (\bibinfo {year} {1982})}\BibitemShut
  {NoStop}%
\bibitem [{\citenamefont {Turner}(1990)}]{PhysRevD.42.3572}%
  \BibitemOpen
  \bibfield  {author} {\bibinfo {author} {\bibfnamefont {M.}~\bibnamefont
  {Turner}},\ }\href {\doibase 10.1103/PhysRevD.42.3572} {\bibfield  {journal}
  {\bibinfo  {journal} {Phys. Rev. D}\ }\textbf {\bibinfo {volume} {42}},\
  \bibinfo {pages} {3572} (\bibinfo {year} {1990})}\BibitemShut {NoStop}%
\bibitem [{\citenamefont {Chou}\ \emph {et~al.}(2008)\citenamefont {Chou},
  \citenamefont {Wester}, \citenamefont {Baumbaugh}, \citenamefont {Gustafson},
  \citenamefont {Irizarry-Valle}, \citenamefont {Mazur}, \citenamefont
  {Steffen}, \citenamefont {Tomlin}, \citenamefont {Yang},\ and\ \citenamefont
  {Yoo}}]{PhysRevLett.100.080402}%
  \BibitemOpen
  \bibfield  {author} {\bibinfo {author} {\bibfnamefont {A.~S.}\ \bibnamefont
  {Chou}}, \bibinfo {author} {\bibfnamefont {W.}~\bibnamefont {Wester}},
  \bibinfo {author} {\bibfnamefont {A.}~\bibnamefont {Baumbaugh}}, \bibinfo
  {author} {\bibfnamefont {H.~R.}\ \bibnamefont {Gustafson}}, \bibinfo {author}
  {\bibfnamefont {Y.}~\bibnamefont {Irizarry-Valle}}, \bibinfo {author}
  {\bibfnamefont {P.~O.}\ \bibnamefont {Mazur}}, \bibinfo {author}
  {\bibfnamefont {J.~H.}\ \bibnamefont {Steffen}}, \bibinfo {author}
  {\bibfnamefont {R.}~\bibnamefont {Tomlin}}, \bibinfo {author} {\bibfnamefont
  {X.}~\bibnamefont {Yang}}, \ and\ \bibinfo {author} {\bibfnamefont
  {J.}~\bibnamefont {Yoo}},\ }\href {\doibase 10.1103/PhysRevLett.100.080402}
  {\bibfield  {journal} {\bibinfo  {journal} {Phys. Rev. Lett.}\ }\textbf
  {\bibinfo {volume} {100}},\ \bibinfo {pages} {080402} (\bibinfo {year}
  {2008})}\BibitemShut {NoStop}%
\bibitem [{\citenamefont {Ehret}\ \emph {et~al.}(2010)\citenamefont {Ehret},
  \citenamefont {Frede}, \citenamefont {Ghazaryan}, \citenamefont
  {Hildebrandt}, \citenamefont {Knabbe}, \citenamefont {Kracht}, \citenamefont
  {Lindner}, \citenamefont {List}, \citenamefont {Meier}, \citenamefont
  {Meyer}, \citenamefont {Notz}, \citenamefont {Redondo}, \citenamefont
  {Ringwald}, \citenamefont {Wiedemann},\ and\ \citenamefont
  {Willke}}]{Ehret2010149}%
  \BibitemOpen
  \bibfield  {author} {\bibinfo {author} {\bibfnamefont {K.}~\bibnamefont
  {Ehret}}, \bibinfo {author} {\bibfnamefont {M.}~\bibnamefont {Frede}},
  \bibinfo {author} {\bibfnamefont {S.}~\bibnamefont {Ghazaryan}}, \bibinfo
  {author} {\bibfnamefont {M.}~\bibnamefont {Hildebrandt}}, \bibinfo {author}
  {\bibfnamefont {E.-A.}\ \bibnamefont {Knabbe}}, \bibinfo {author}
  {\bibfnamefont {D.}~\bibnamefont {Kracht}}, \bibinfo {author} {\bibfnamefont
  {A.}~\bibnamefont {Lindner}}, \bibinfo {author} {\bibfnamefont
  {J.}~\bibnamefont {List}}, \bibinfo {author} {\bibfnamefont {T.}~\bibnamefont
  {Meier}}, \bibinfo {author} {\bibfnamefont {N.}~\bibnamefont {Meyer}},
  \bibinfo {author} {\bibfnamefont {D.}~\bibnamefont {Notz}}, \bibinfo {author}
  {\bibfnamefont {J.}~\bibnamefont {Redondo}}, \bibinfo {author} {\bibfnamefont
  {A.}~\bibnamefont {Ringwald}}, \bibinfo {author} {\bibfnamefont
  {G.}~\bibnamefont {Wiedemann}}, \ and\ \bibinfo {author} {\bibfnamefont
  {B.}~\bibnamefont {Willke}},\ }\href {\doibase
  http://dx.doi.org/10.1016/j.physletb.2010.04.066} {\bibfield  {journal}
  {\bibinfo  {journal} {Physics Letters B}\ }\textbf {\bibinfo {volume}
  {689}},\ \bibinfo {pages} {149 } (\bibinfo {year} {2010})}\BibitemShut
  {NoStop}%
\bibitem [{\citenamefont {Arik}\ \emph {et~al.}(2011)\citenamefont {Arik} \emph
  {et~al.}}]{PhysRevLett.107.261302}%
  \BibitemOpen
  \bibfield  {author} {\bibinfo {author} {\bibfnamefont {M.}~\bibnamefont
  {Arik}} \emph {et~al.} (\bibinfo {collaboration} {CAST Collaboration}),\
  }\href {\doibase 10.1103/PhysRevLett.107.261302} {\bibfield  {journal}
  {\bibinfo  {journal} {Phys. Rev. Lett.}\ }\textbf {\bibinfo {volume} {107}},\
  \bibinfo {pages} {261302} (\bibinfo {year} {2011})}\BibitemShut {NoStop}%
\bibitem [{\citenamefont {Kampwirth}\ \emph {et~al.}(1977)\citenamefont
  {Kampwirth}, \citenamefont {Hafstrom},\ and\ \citenamefont {Wu}}]{Nb3Sn}%
  \BibitemOpen
  \bibfield  {author} {\bibinfo {author} {\bibfnamefont {R.}~\bibnamefont
  {Kampwirth}}, \bibinfo {author} {\bibfnamefont {J.}~\bibnamefont {Hafstrom}},
  \ and\ \bibinfo {author} {\bibfnamefont {C.}~\bibnamefont {Wu}},\ }\href
  {http://ieeexplore.ieee.org/xpls/abs_all.jsp?arnumber=1059445} {\bibfield
  {journal} {\bibinfo  {journal} {Magnetics, IEEE Transactions on}\ }\textbf
  {\bibinfo {volume} {13}},\ \bibinfo {pages} {315} (\bibinfo {year}
  {1977})}\BibitemShut {NoStop}%
\bibitem [{\citenamefont {Kuhr}\ \emph {et~al.}(2007)\citenamefont {Kuhr},
  \citenamefont {Gleyzes}, \citenamefont {Guerlin}, \citenamefont {Bernu},
  \citenamefont {Hoff}, \citenamefont {Deléglise}, \citenamefont {Osnaghi},
  \citenamefont {Brune}, \citenamefont {Raimond}, \citenamefont {Haroche} \emph
  {et~al.}}]{:/content/aip/journal/apl/90/16/10.1063/1.2724816}%
  \BibitemOpen
  \bibfield  {author} {\bibinfo {author} {\bibfnamefont {S.}~\bibnamefont
  {Kuhr}}, \bibinfo {author} {\bibfnamefont {S.}~\bibnamefont {Gleyzes}},
  \bibinfo {author} {\bibfnamefont {C.}~\bibnamefont {Guerlin}}, \bibinfo
  {author} {\bibfnamefont {J.}~\bibnamefont {Bernu}}, \bibinfo {author}
  {\bibfnamefont {U.~B.}\ \bibnamefont {Hoff}}, \bibinfo {author}
  {\bibfnamefont {S.}~\bibnamefont {Deléglise}}, \bibinfo {author}
  {\bibfnamefont {S.}~\bibnamefont {Osnaghi}}, \bibinfo {author} {\bibfnamefont
  {M.}~\bibnamefont {Brune}}, \bibinfo {author} {\bibfnamefont {J.-M.}\
  \bibnamefont {Raimond}}, \bibinfo {author} {\bibfnamefont {S.}~\bibnamefont
  {Haroche}},  \emph {et~al.},\ }\href {\doibase
  http://dx.doi.org/10.1063/1.2724816} {\bibfield  {journal} {\bibinfo
  {journal} {Applied Physics Letters}\ }\textbf {\bibinfo {volume} {90}},\
  \bibinfo {eid} {164101} (\bibinfo {year} {2007})}\BibitemShut {NoStop}%
\bibitem [{\citenamefont {Yurke}\ \emph {et~al.}(1996)\citenamefont {Yurke},
  \citenamefont {Roukes}, \citenamefont {Movshovich},\ and\ \citenamefont
  {Pargellis}}]{Yurke:1996dt}%
  \BibitemOpen
  \bibfield  {author} {\bibinfo {author} {\bibfnamefont {B.}~\bibnamefont
  {Yurke}}, \bibinfo {author} {\bibfnamefont {M.~L.}\ \bibnamefont {Roukes}},
  \bibinfo {author} {\bibfnamefont {R.}~\bibnamefont {Movshovich}}, \ and\
  \bibinfo {author} {\bibfnamefont {A.~N.}\ \bibnamefont {Pargellis}},\ }\href
  {\doibase 10.1063/1.116845} {\bibfield  {journal} {\bibinfo  {journal}
  {Applied Physics Letters}\ }\textbf {\bibinfo {volume} {69}},\ \bibinfo
  {pages} {3078} (\bibinfo {year} {1996})}\BibitemShut {NoStop}%
\bibitem [{\citenamefont {Chen}\ \emph {et~al.}(2011)\citenamefont {Chen},
  \citenamefont {Hover}, \citenamefont {Sendelbach}, \citenamefont {Maurer},
  \citenamefont {Merkel}, \citenamefont {Pritchett}, \citenamefont {Wilhelm},\
  and\ \citenamefont {McDermott}}]{Chen:2011fj}%
  \BibitemOpen
  \bibfield  {author} {\bibinfo {author} {\bibfnamefont {Y.~F.}\ \bibnamefont
  {Chen}}, \bibinfo {author} {\bibfnamefont {D.}~\bibnamefont {Hover}},
  \bibinfo {author} {\bibfnamefont {S.}~\bibnamefont {Sendelbach}}, \bibinfo
  {author} {\bibfnamefont {L.}~\bibnamefont {Maurer}}, \bibinfo {author}
  {\bibfnamefont {S.~T.}\ \bibnamefont {Merkel}}, \bibinfo {author}
  {\bibfnamefont {E.~J.}\ \bibnamefont {Pritchett}}, \bibinfo {author}
  {\bibfnamefont {F.~K.}\ \bibnamefont {Wilhelm}}, \ and\ \bibinfo {author}
  {\bibfnamefont {R.}~\bibnamefont {McDermott}},\ }\href {\doibase
  10.1103/PhysRevLett.107.217401} {\bibfield  {journal} {\bibinfo  {journal}
  {Physical Review Letters}\ }\textbf {\bibinfo {volume} {107}},\ \bibinfo
  {pages} {217401} (\bibinfo {year} {2011})}\BibitemShut {NoStop}%
\bibitem [{\citenamefont {Lamoreaux}\ \emph {et~al.}(2013)\citenamefont
  {Lamoreaux}, \citenamefont {van Bibber}, \citenamefont {Lehnert},\ and\
  \citenamefont {Carosi}}]{PhysRevD.88.035020}%
  \BibitemOpen
  \bibfield  {author} {\bibinfo {author} {\bibfnamefont {S.}~\bibnamefont
  {Lamoreaux}}, \bibinfo {author} {\bibfnamefont {K.}~\bibnamefont {van
  Bibber}}, \bibinfo {author} {\bibfnamefont {K.}~\bibnamefont {Lehnert}}, \
  and\ \bibinfo {author} {\bibfnamefont {G.}~\bibnamefont {Carosi}},\ }\href
  {\doibase 10.1103/PhysRevD.88.035020} {\bibfield  {journal} {\bibinfo
  {journal} {Phys. Rev. D}\ }\textbf {\bibinfo {volume} {88}},\ \bibinfo
  {pages} {035020} (\bibinfo {year} {2013})}\BibitemShut {NoStop}%
\bibitem [{\citenamefont {DePanfilis}\ \emph {et~al.}(1987)\citenamefont
  {DePanfilis}, \citenamefont {Melissinos}, \citenamefont {Moskowitz},
  \citenamefont {Rogers}, \citenamefont {Semertzidis}, \citenamefont {Wuensch},
  \citenamefont {Halama}, \citenamefont {Prodell}, \citenamefont {Fowler},\
  and\ \citenamefont {Nezrick}}]{PhysRevLett.59.839}%
  \BibitemOpen
  \bibfield  {author} {\bibinfo {author} {\bibfnamefont {S.}~\bibnamefont
  {DePanfilis}}, \bibinfo {author} {\bibfnamefont {A.~C.}\ \bibnamefont
  {Melissinos}}, \bibinfo {author} {\bibfnamefont {B.~E.}\ \bibnamefont
  {Moskowitz}}, \bibinfo {author} {\bibfnamefont {J.~T.}\ \bibnamefont
  {Rogers}}, \bibinfo {author} {\bibfnamefont {Y.~K.}\ \bibnamefont
  {Semertzidis}}, \bibinfo {author} {\bibfnamefont {W.~U.}\ \bibnamefont
  {Wuensch}}, \bibinfo {author} {\bibfnamefont {H.~J.}\ \bibnamefont {Halama}},
  \bibinfo {author} {\bibfnamefont {A.~G.}\ \bibnamefont {Prodell}}, \bibinfo
  {author} {\bibfnamefont {W.~B.}\ \bibnamefont {Fowler}}, \ and\ \bibinfo
  {author} {\bibfnamefont {F.~A.}\ \bibnamefont {Nezrick}},\ }\href {\doibase
  10.1103/PhysRevLett.59.839} {\bibfield  {journal} {\bibinfo  {journal} {Phys.
  Rev. Lett.}\ }\textbf {\bibinfo {volume} {59}},\ \bibinfo {pages} {839}
  (\bibinfo {year} {1987})}\BibitemShut {NoStop}%
\bibitem [{\citenamefont {Wuensch}\ \emph {et~al.}(1989)\citenamefont
  {Wuensch}, \citenamefont {De~Panfilis-Wuensch}, \citenamefont {Semertzidis},
  \citenamefont {Rogers}, \citenamefont {Melissinos}, \citenamefont {Halama},
  \citenamefont {Moskowitz}, \citenamefont {Prodell}, \citenamefont {Fowler},\
  and\ \citenamefont {Nezrick}}]{PhysRevD.40.3153}%
  \BibitemOpen
  \bibfield  {author} {\bibinfo {author} {\bibfnamefont {W.~U.}\ \bibnamefont
  {Wuensch}}, \bibinfo {author} {\bibfnamefont {S.}~\bibnamefont
  {De~Panfilis-Wuensch}}, \bibinfo {author} {\bibfnamefont {Y.~K.}\
  \bibnamefont {Semertzidis}}, \bibinfo {author} {\bibfnamefont {J.~T.}\
  \bibnamefont {Rogers}}, \bibinfo {author} {\bibfnamefont {A.~C.}\
  \bibnamefont {Melissinos}}, \bibinfo {author} {\bibfnamefont {H.~J.}\
  \bibnamefont {Halama}}, \bibinfo {author} {\bibfnamefont {B.~E.}\
  \bibnamefont {Moskowitz}}, \bibinfo {author} {\bibfnamefont {A.~G.}\
  \bibnamefont {Prodell}}, \bibinfo {author} {\bibfnamefont {W.~B.}\
  \bibnamefont {Fowler}}, \ and\ \bibinfo {author} {\bibfnamefont {F.~A.}\
  \bibnamefont {Nezrick}},\ }\href {\doibase 10.1103/PhysRevD.40.3153}
  {\bibfield  {journal} {\bibinfo  {journal} {Phys. Rev. D}\ }\textbf {\bibinfo
  {volume} {40}},\ \bibinfo {pages} {3153} (\bibinfo {year}
  {1989})}\BibitemShut {NoStop}%
\bibitem [{\citenamefont {Hagmann}\ \emph {et~al.}(1990)\citenamefont
  {Hagmann}, \citenamefont {Sikivie}, \citenamefont {Sullivan},\ and\
  \citenamefont {Tanner}}]{PhysRevD.42.1297}%
  \BibitemOpen
  \bibfield  {author} {\bibinfo {author} {\bibfnamefont {C.}~\bibnamefont
  {Hagmann}}, \bibinfo {author} {\bibfnamefont {P.}~\bibnamefont {Sikivie}},
  \bibinfo {author} {\bibfnamefont {N.~S.}\ \bibnamefont {Sullivan}}, \ and\
  \bibinfo {author} {\bibfnamefont {D.~B.}\ \bibnamefont {Tanner}},\ }\href
  {\doibase 10.1103/PhysRevD.42.1297} {\bibfield  {journal} {\bibinfo
  {journal} {Phys. Rev. D}\ }\textbf {\bibinfo {volume} {42}},\ \bibinfo
  {pages} {1297} (\bibinfo {year} {1990})}\BibitemShut {NoStop}%
\bibitem [{\citenamefont {Tanner}(2013)}]{admxpatras2013}%
  \BibitemOpen
  \bibfield  {author} {\bibinfo {author} {\bibfnamefont {D.}~\bibnamefont
  {Tanner}},\ }in\ \href@noop {} {\emph {\bibinfo {booktitle} {Proceedings of
  the 9th Patras Workshop on Axions, WIMPS and WISPs}}},\ \bibinfo {editor}
  {edited by\ \bibinfo {editor} {\bibfnamefont {U.}~\bibnamefont {Oberlack}}\
  and\ \bibinfo {editor} {\bibfnamefont {P.}~\bibnamefont {Sissol}}}\ (\bibinfo
  {year} {2013})\ pp.\ \bibinfo {pages} {171 -- 176}\BibitemShut {NoStop}%
\end{thebibliography}%

\end{document}